\documentstyle[preprint,aps,psfig]{revtex}

\begin{document}
\draft

\title{ Reply to Comment on "Criterion that Determines the Foldability
of Proteins"}
\author{D. K. Klimov and D. Thirumalai}
\address{Institute for Physical Science and Technology and
Department of Chemistry and Biochemistry\\
University of Maryland, College Park, Maryland 20742}
\maketitle

\begin{abstract}

We point out that the correlation between folding times and $\sigma =
(T_{\theta } - T_{f})/T_{\theta }$ in protein-like heteropolymer models
where $T_{\theta }$ and $T_{f}$ are the collapse and folding transition
temperatures was already established in 1993 \textit{ before the other
presumed equivalent } criterion (folding times correlating with $T_{f}$
alone) was suggested.  We argue that the folding times for these models
show no useful correlation with the energy gap even if restricted to the
ensemble of compact structures as suggested by Karplus and Shakhnovich. 

\end{abstract}
\vspace{1cm}

In a recent article \cite{KT} we (KT) showed that for certain lattice
models of proteins the folding times, \(\tau _{f}\), correlate extremely
well with \(\sigma =(T_{\theta } -T_{f})/T_{\theta }\) where \(T_{\theta
}\) and \(T_{f}\) 
are well defined thermodynamic collapse and folding transition
temperatures, respectively. We also demonstrated that there is "no useful
correlation between folding times and the energy gap between the native
conformation and the first excited state" \cite{KT}. In response to these
results Karplus and Shakhnovich (KS) \cite{KS} try to argue (i) that the
folding criterion used by KT is "essentially the same as one introduced
earlier  \cite{Sali94b,Abk2}" and (ii) that the energy gap used by KT is not
"appropriate". We take up these two issues separately. 
In addition, 
we also show that folding times do not correlate with the energy gap
\(\Delta _{CS}\) restricted to the ensemble of compact structures as KS
\cite{KS} desire. 

First we settle the historical claims to priority. The criterion used by KT
was {\em already established in 1993} in \cite{Cam93} which was 
\textit{published before} \cite{Abk2} was even received for review. 
There is no plot of folding time versus \(T_{f}\) in \cite{Sali94b},
an article that was submitted \textit{after} \cite{Cam93} was accepted for
publication. 
It was shown in
\cite{Cam93} (cited in \cite{Abk2}) that folding times correlate with
\(\sigma \).  We showed using numerical results and theoretical arguments
that "there appears to be useful correlation between folding time \(\tau
_{r}\) and \(\sigma =1-T_{f}/T_{\theta }\): the smaller the value of
\(\sigma \) the smaller the value of \(\tau _{r}\)" \cite{Cam93}. 

Having addressed the historical claim to priority we now examine the
statement that our folding criterion is "essentially the same as one
introduced earlier" \cite{Abk2}. This claim is based on the
observation that there is a correlation between folding time and \(T_{f}\)
seen in "right most part of Fig. (8a) of \cite{Abk2}". Fig. (8a) of
\cite{Abk2} shows the folding time for 10 sequences of which 5 lie in the
"right most part". The simulation temperature for the data in Fig. (8a) is
\(T_{s} = 1.0\) and the \(T_{f}\) for all the sequences lie in the range
\(0.63 \leq T_{f} \leq 1.06\) (see Table I in \cite{Abk2}). Of the 5
sequences that lie in the "right most part"  of Fig. (8a) only one has
\(T_{f} > T_{s}\). The apparent correlation (for the five sequences
for which the folding times \textit{merely change} by a factor of 3) 
claimed by KS occurs when the
native state is not stable. In Fig. (8b) in \cite{Abk2} a plot of
folding time as a function of \(T_{f}\) at \(T_{s} = 0.7\) (which is below
\(T_{f}\) for all sequences except two) is shown. In this figure, with
\(T_{f} > T_{s} \), one observes \textit{no correlation whatsoever}
between \(\tau _{f}\) and \(T_{f}\). In contrast, our results \cite{KT}
show that \(\tau _{f}\) (which spans six orders of magnitude) correlates well with \(\sigma \).  It is the
\textit{interplay between the two intrinsic sequence dependent temperatures
\(T_{\theta }\) and \(T_{f}\)} that seems to correlate with \(\tau _{f}\)
under conditions when the native state has the highest occupation
probability. 

It was forcefully stated in \cite{Sali94a,Karp95} that the "necessary
and sufficient"  condition for folding in these models is that there
should be "a pronounced energy gap between the native and first
excited state for the fully compact ensemble" \cite{Karp95}. The
reason we did not display \(\tau _{f}\) as a function of \(\Delta
_{CS}\) in \cite{KT} is that roughly  half of the sequences had
non-compact native conformations.  It is, therefore,  not clear why one
should be restricted only to the ensemble of compact structures.  However,
the folding times for sequences in \cite{KT} when plotted as a
function of \(\Delta _{CS}\), as KS desire, also does not show any  
correlation (see Fig. (1a)). There are a large number of sequences, all
with roughly the same folding times, but with very different values of
\(\Delta _{CS}\). The lack of correlation between the energy gap and
foldability has been established  for other models as well
\cite{Unger,Fukugita,Camacho96}. 

There are a few additional comments that are worth making. (1) KS claim that
\(\tau _{f}\) should correlate well with both \(T_{f}\) and \(\Delta
_{CS}\). This implies that \(T_{f}\) should correlate with  
\(\Delta _{CS}\). In
Fig. (1b) we plot \(T_{f}\) as a function of \(\Delta _{CS}\). The lack of
correlation is evident. (2) Unlike \(T_{\theta }\) and \(T_{f}\), 
\(\Delta _{CS}\) is not experimentally
measurable. Moreover, with the exception of two studies 
\cite{Fukugita,Honey90}, the
energy gap is not easily computable for simple off-lattice
models. Thus the
practical utility of \(\Delta _{CS}\) is not clear. (3) We should
emphasize that all the criteria proposed for folding times should be
viewed as statistical. This means that, under conditions when the native
state is significantly populated, \(\tau _{f}\) may correlate with
certain properties intrinsic to the sequence provided a number of
sequences is studied. Our results suggest that \(\sigma \) could be one
of such properties.

\begin{figure}

{\bf Fig. 1.} (a) Folding time \(\tau _{f}\) versus 
\(\Delta _{CS}\). 
(b) \(T_{f}\) as a 
function of \(\Delta
_{CS}\).  
Solid circles correspond to 15-mer and open circles are for  27-mer. 
Right and upper axes are for 27-mer.

\end{figure}

\newpage

\begin{center}
\begin{minipage}{15cm}
\[
\psfig{figure=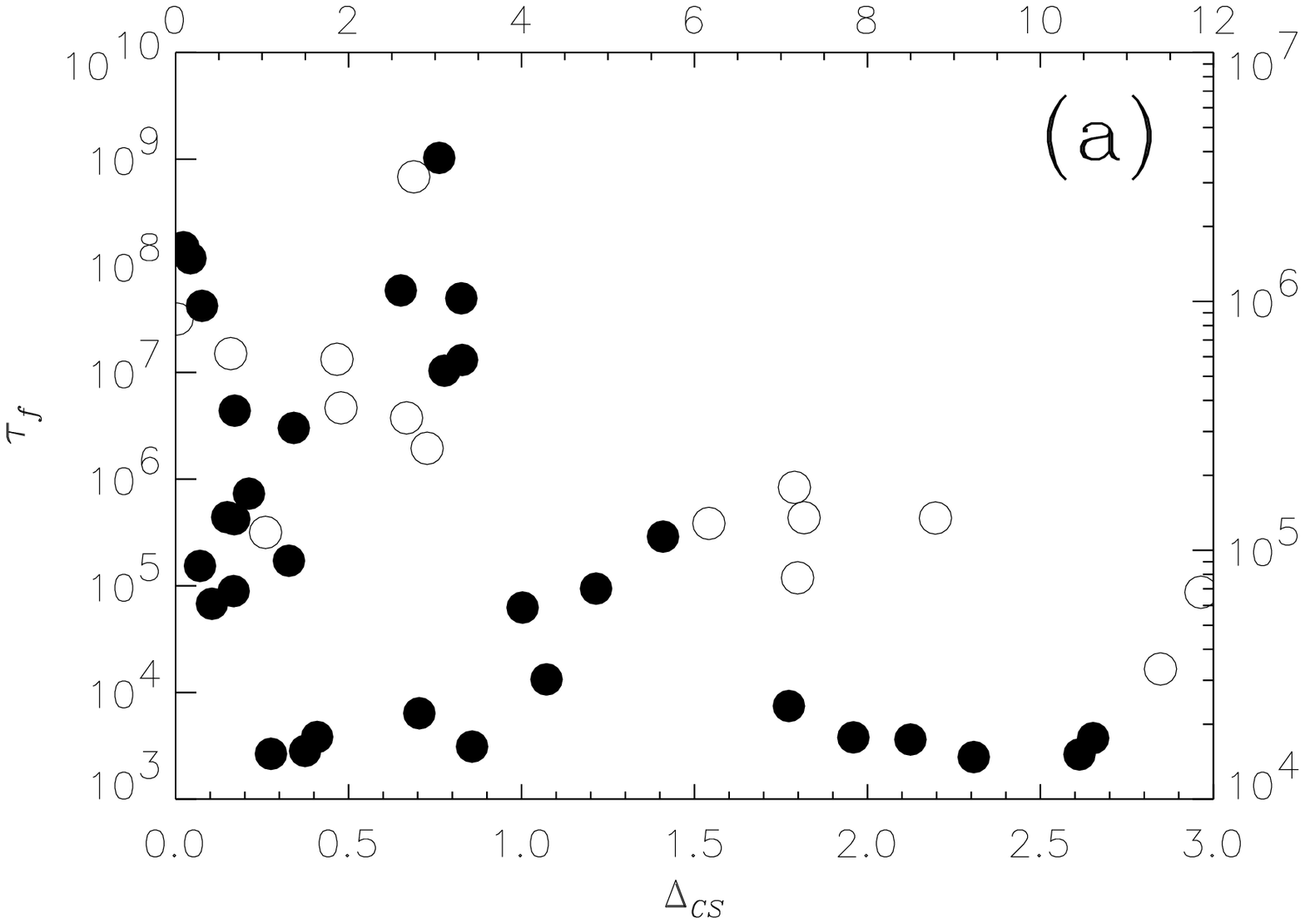,height=10cm,width=14cm}
\]
\end{minipage}

\end{center}  

\begin{center}
\begin{minipage}{15cm}
\[
\psfig{figure=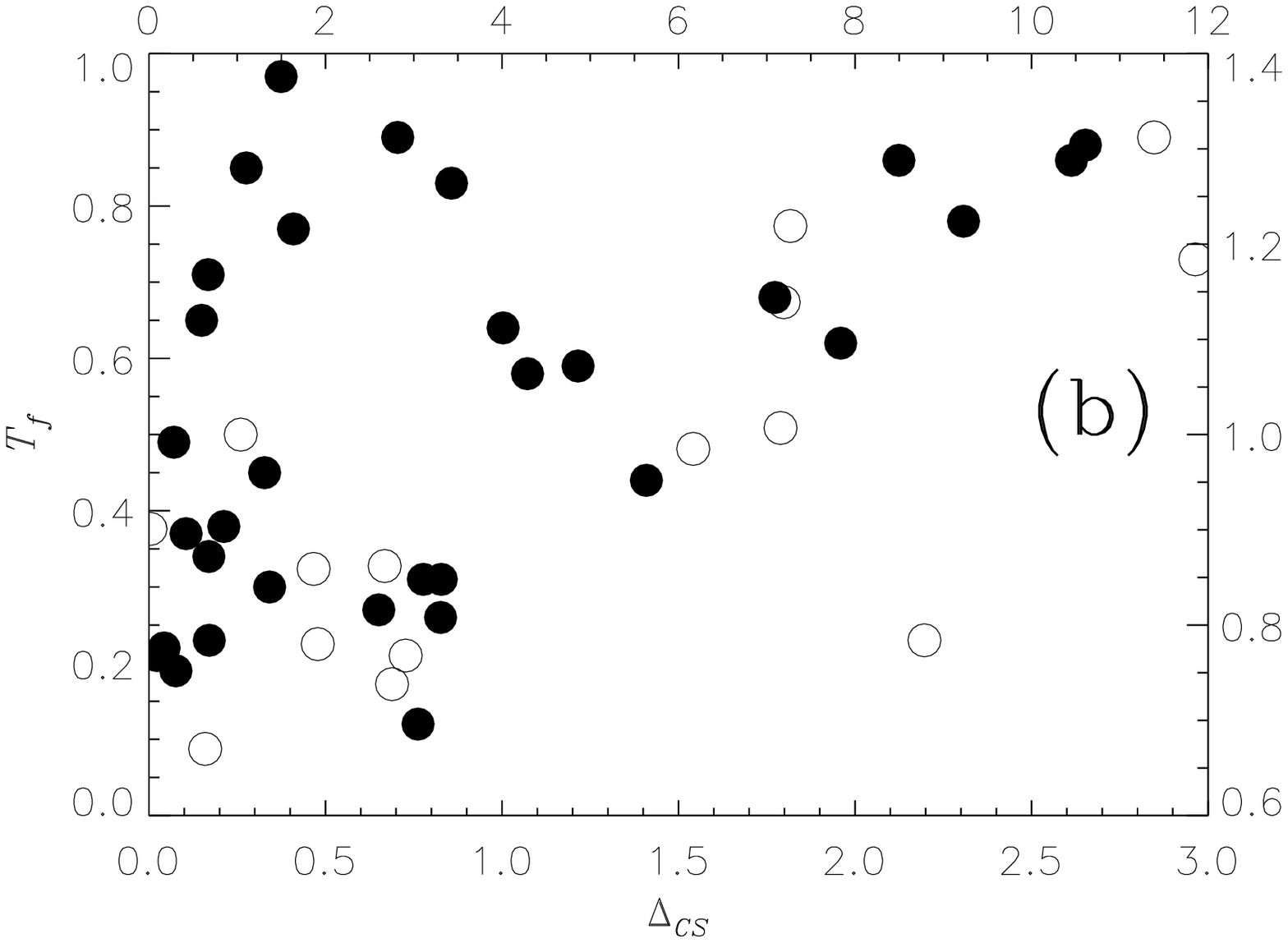,height=10cm,width=14cm}
\]
\end{minipage}

{\bf \large Fig. 1}\\
\end{center}

\end{document}